\documentclass[twocolumn,fp]{jpsj3}

\usepackage{txfonts}

\usepackage{amsmath}
\usepackage{amsfonts}
\usepackage{bm}
\usepackage{graphicx}
\usepackage{color}

\title{
Method to solve quantum few-body problems with artificial neural networks
}

\author{Hiroki Saito}
\inst{Department of Engineering Science, University of
Electro-Communications, Tokyo 182-8585, Japan}

\abst{
A machine learning technique to obtain the ground states of quantum few-body
systems using artificial neural networks is developed.
Bosons in continuous space are considered and a neural network is optimized
in such a way that when particle positions are input into the network,
the ground-state wave function is output from the network.
The method is applied to the Calogero-Sutherland model in one-dimensional
space and Efimov bound states in three-dimensional space.
}

\begin{document}
\maketitle

\section{Introduction}

Quantum many-body problems are difficult because the size of the Hilbert
space increases exponentially with the number of degrees of freedom.
For example, in quantum spin problems, each spin state is represented by two
bases, i.e., spin-up and spin-down states, and hence the total number of
bases to represent the quantum state of $N$ spins is $2^N$.
In numerical calculations, therefore, the amount of memory required to store
the quantum states and the computational time required to handle them
increases exponentially with $N$.
This hinders precise numerical analysis of quantum many-body systems with
large $N$.
To circumvent this problem, various methods have been developed, such as
quantum Monte Carlo method~\cite{Foulkes}, the density matrix
renormalization group~\cite{Scholl}, and tensor networks~\cite{Ran}.

Recently, a method to solve quantum many-body problems using artificial
neural networks was proposed~\cite{Carleo}.
In this work, it was shown that quantum many-body states can be efficiently
stored in an artificial neural network~\cite{Gao,Torlai08,Cai,Schmitt},
called a restricted Boltzmann machine, where the number of parameters used
to represent the neural network is much smaller than the size of the Hilbert
space.
Using this neural network representation, it was demonstrated that
the ground state and time evolution of quantum spin systems can be
obtained~\cite{Carleo,Czischek}.
This method has subsequently been applied to the Bose-Hubbard
model~\cite{SaitoL} using feedforward neural networks and to the
Fermi-Hubbard model~\cite{Nomura}.
Deep neural networks with multiple hidden layers can efficiently represent
quantum many-body states~\cite{Saito,Carleo08}.
Quantum many-body states in neural networks have also been studied from the
perspective of tensor networks~\cite{YHuang,Chen,Glasser,You} and quantum
entanglement~\cite{DengX,Deng}.

All of the above-mentioned studies and other studies on applications of
neural networks to physics~\cite{Carra,Nieu,Ohtsuki,Tanaka}
have considered spatially discrete systems, i.e., spins or particles on
lattices.
In the present paper, by contrast, quantum many-body problems in continuous
space are solved using artificial neural networks.
We consider interacting several bosons, and the positions of these particles
$x_1$, $x_2$, $\cdots, x_N$ or their appropriate functions are input into
the neural network, which is to be optimized to output the ground-state wave
function $\psi(x_1, x_2, \cdots, x_N)$.
This contrasts with previous studies, in which discrete variables, such as
spin states and particle numbers on lattices, are input into neural
networks.
It is well known that any continuous function can be represented by a neural
network~\cite{onlinebook}, and in fact, it was demonstrated that a
single-particle Schr\"odinger equation can be solved using a neural
network~\cite{Teng}.
However, also in the case of continuous space, the size of the Hilbert
space increases exponentially with the number of particles, and it is not
obvious that such wave functions with large degrees of freedom can be
efficiently represented by neural networks.

In the present paper, we will show that the ground-state wave functions of
many-body problems in continuous space can be obtained by a method using
artificial neural networks.
The positions of particles are transformed appropriately and input into the
neural network, which is optimized to output the ground-state wave
function.
We apply our method to the Calogero-Sutherland model in one-dimensional
space and the obtained wave functions are shown to be close to the exact
solution of the ground state.
The method is also applied to bosons with resonant interaction in
three-dimensional space, and the Efimov bound states are obtained.

The remainder of the paper is organized as follows.
Section~\ref{s:method} describes a method to obtain the wave function of
the many-body ground state using neural networks.
Section~\ref{s:results} presents numerical results for one-dimensional and
three-dimensional problems.
Finally, Section~\ref{s:conc} offers conclusions.

\section{Method}
\label{s:method}

\begin{figure}[tb]
\includegraphics[width=8cm]{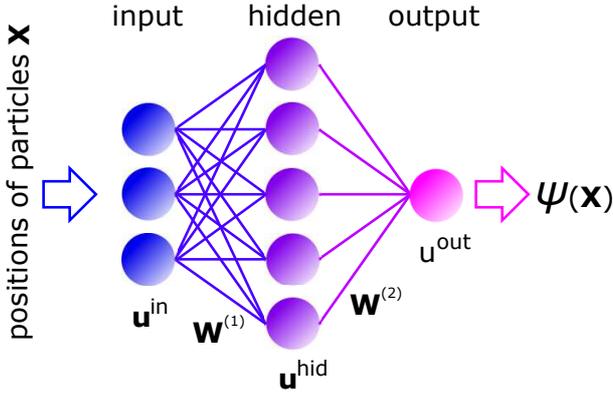}
\caption{
(Color online) Schematic illustration of the fully connected feedforward
neural network to obtain the few-body ground states.
The positions of particles $\bm{X}$ or their functions are input into the
network.
Units in the hidden layer are connected to all input units and the output
unit.
The output $u^{\rm out}$ of the network gives the wave function as
$\psi(\bm{X}) = \exp(u^{\rm out})$.
}
\label{f:schematic}
\end{figure}
We use a feedforward artificial neural network, as illustrated in
Fig.~\ref{f:schematic}.
The input units $u_1^{\rm in}, u_2^{\rm in}, \cdots, u_{N_{\rm in}}^{\rm
in}$ are set to real numbers, where $N_{\rm in}$ is the number of input
units.
The value of each hidden unit is generated by all the input units through
weights as
\begin{equation}
u_i^{\rm hid} = \sum_{j=1}^{N_{\rm in}} W_{ji}^{(1)} u_j^{\rm in} + b_i^{(1)},
\end{equation}
where the weights $W_{ji}^{(1)}$ and biases $b_i^{(1)}$ are real numbers.
In our feedforward network, there is only a single output, whose value is
given by
\begin{equation}
u^{\rm out} = \sum_{i=1}^{N_{\rm hid}} W_i^{(2)} f(u_i^{\rm hid}),
\end{equation}
where $N_{\rm hid}$ is the number of hidden units, the weights $W_i^{(2)}$
are real numbers, and $f$ is the activation function.
In this paper, we adopt the hyperbolic tangent function, $f(x) = \tanh(x)$,
as the activation function.
We thus use a fully connected network with a single hidden layer.
The number of parameters $W_{ji}^{(1)}$, $b_i^{(1)}$, and $W_i^{(2)}$ in
this neural network is $(N_{\rm in} + 2) N_{\rm hid}$.

We consider a system of $N$ identical bosons.
Particle positions are written as $\bm{X} = (x_1, x_2, \cdots, x_N)$
for one dimension and $\bm{X} = (\bm{x}_1, \bm{x}_2, \cdots, \bm{x}_N)$ for
higher dimensions.
We input $\bm{\xi}(\bm{X}) = (\xi_1(\bm{X}), \xi_2(\bm{X}), \cdots,
\xi_{N_{\rm in}}(\bm{X}))$ into the neural network, where
$\xi_i(\bm{X})$ are functions of the particle positions.
The functions $\xi_i(\bm{X})$ are chosen to facilitate the representation of
many-body states in the neural network, and depend on the problem, which
will be specified later.
The many-body wave function is given by the output unit as
\begin{equation}
  \psi(\bm{X}) = \exp(u^{\rm out}).
\end{equation}
This wave function is not normalized.
Our aim is to optimize the network parameters in such a way that if we input
$\bm{\xi}(\bm{X})$ into the neural network, the output $\psi(\bm{X}) =
\exp(u^{\rm out})$ provides an approximate ground-state wave function.
In other words, we use neural networks as variational wave functions.
Since the ground-state wave functions of bosons can be taken to be positive
everywhere, we only use a single output unit that gives a positive wave
function $\psi(\bm{X}) = \exp(u^{\rm out})$.
If we use a network with two output units, complex-valued wave functions can
be represented~\cite{SaitoL,Saito}.

The expectation value of a quantity $\hat A$ is calculated as follows.
To deal with a large Hilbert space, we calculate the integration $\int
d\bm{X}$ using the Monte Carlo method.
Through Metropolis sampling, we obtain a series of samples $(\bm{X}_1,
\bm{X}_2, \cdots, \bm{X}_{N_{\rm sample}})$, where the probability
$P(\bm{X})$ of $\bm{X}$ being sampled is proportional to $\psi^2(\bm{X})$.
Using these samples, the expectation value is calculated as
\begin{eqnarray}
\langle \hat A \rangle & = & \frac{\int \psi(\bm{X}) \hat A \psi(\bm{X})
  d\bm{X}}{\int \psi^2(\bm{X}) d\bm{X}} \nonumber \\
& = & \int P(\bm{X}) \psi^{-1}(\bm{X}) \hat A \psi(\bm{X}) d\bm{X}
\nonumber \\
& \simeq & \frac{1}{N_{\rm sample}} \sum_{i=1}^{N_{\rm sample}}
\psi^{-1}(\bm{X}_i) \hat A \psi(\bm{X}_i) \nonumber \\
& \equiv & \langle \psi^{-1} \hat A \psi \rangle_s.
\end{eqnarray}
To minimize the energy $E$ of the system, we need to calculate the
derivative of energy with respect to the network parameters as
\begin{eqnarray} \label{grad}
\frac{\partial E}{\partial W} & = & \frac{\partial}{\partial W}
\frac{\int \psi(\bm{X}) \hat H \psi(\bm{X}) d\bm{X}}{\int \psi^2(\bm{X})
  d\bm{X}} \nonumber \\
& \simeq & 2 \langle \psi^{-1} O_W \hat H \psi \rangle_s
- 2 \langle \psi^{-1} \hat H \psi \rangle_s \langle O_W \rangle_s,
\end{eqnarray}
where $W$ is one of the network parameters ($W_{ji}^{(1)}$, $b_i^{(1)}$, and
$W_i^{(2)}$), $\hat H$ is the Hamiltonian of the system, and $O_W =
\psi^{-1} \partial \psi / \partial W$.
Using the gradient in Eq.~(\ref{grad}), the network parameters are updated
using the Adam scheme~\cite{Kingma}, by which energy converges faster than
the steepest descent method~\cite{Saito}.
Typically, $10^4$-$10^5$ updates are needed for energy convergence
with $N_{\rm sample} \sim 10^4$ samples taken in each update.

It is important to choose initial values of the network parameters because
if we use random numbers for this purpose and start the update procedure,
the network state seriously fluctuates so that the calculation breaks down
in most cases.
This is different from previous cases focusing on lattice
systems~\cite{SaitoL,Saito}, in which random numbers in the initial network
parameters worked well.
To prepare appropriate initial network parameters, we train the network so
that it represents a wave function $\Psi_{\rm train}$.
We train the network to maximize the overlap integral $K$ given by
\begin{equation} \label{I}
K = \frac{\left[ \int \Psi_{\rm train}(\bm{X}) \psi(\bm{X}) d\bm{X}
    \right]^2}{\int \Psi^2_{\rm train}(\bm{X}) d\bm{X}
  \int \psi^2(\bm{X}) d\bm{X}}
\simeq \frac{\langle A \rangle_s^2}{\langle A^2 \rangle_s},
\end{equation}
where $A = \Psi_{\rm train} / \psi$.
This is accomplished by using the gradient of $K$ with respect to network
parameters as
\begin{equation} \label{Igrad}
\frac{\partial K}{\partial W} \simeq 2 K \left( \frac{\langle A O_W
\rangle_s}{\langle A \rangle_s} - \langle O_W \rangle_s \right).
\end{equation}
Using the Adam scheme with Eq.~(\ref{Igrad}), $K$ is maximized and $\psi$
converges toward $\Psi_{\rm train}$.
The initial values of the network parameters in this training process
can be random numbers with an appropriate magnitude~\cite{Xavier}.
For example, $\Psi_{\rm train}$ is chosen to be a noninteracting ground
state, and the interaction is introduced adiabatically.

\begin{figure}[tb]
\includegraphics[width=8cm]{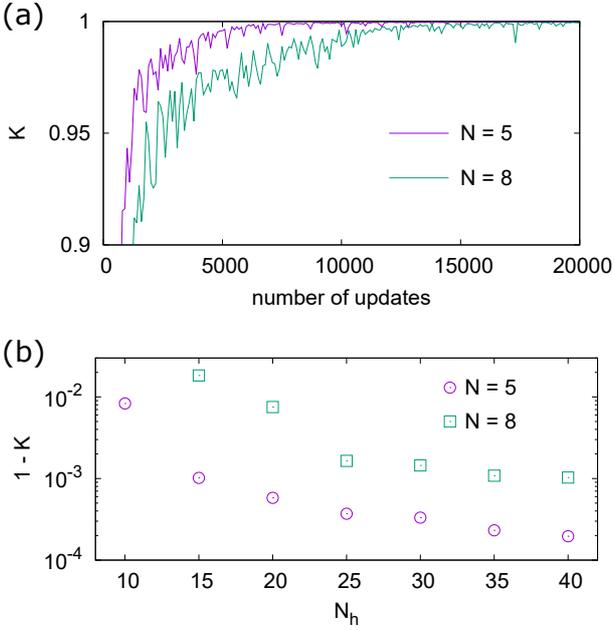}
\caption{
(Color online) Training of the neural network to represent the wave function
in Eq.~(\ref{harmsol}), where the overlap integral $K$ is given in
Eq.~(\ref{I}).
(a) Change in $K$ with respect to the number of network parameter updates
for $N = 5$ and 8.
The number of hidden units is $N_h = 20$.
(b) $N_h$ dependence of $1 - K$ after 20000 updates for $N = 5$ and 8.
The number of samples is $N_{\rm sample} = 10^4$ for each update in (a) and
(b), and $N_{\rm sample} = 10^6$ to calculate the data points in (b).
}
\label{f:train}
\end{figure}
We demonstrate that the foregoing procedure increases $K$ and hence the wave
function $\psi$ that is close to $\Psi_{\rm train}$ can be prepared.
We take the target wave function as
\begin{equation} \label{harmsol}
\Psi_{\rm train}(\bm{X}) = \prod_{i=1}^N e^{-x_i^2 / 2},
\end{equation}
which is the ground state of $N$ bosons in a one-dimensional harmonic
potential.
The neural network inputs are simply the positions of the particles,
\begin{equation} \label{simplexi}
\bm{\xi}(\bm{X}) = (x_1, x_2, \cdots, x_N),
\end{equation}
and therefore, the number of input units is $N_{\rm in} = N$.
Figure~\ref{f:train}(a) shows how $K$ changes in the process of optimizing
the network parameters.
The value of $K$ approaches unity as the update steps proceed, and the
wave function represented by the neural network approaches the target wave
function $\Psi_{\rm train}$.
The finally obtained overlap $K$ decreases with the number of atoms $N$, and
increases with the number of hidden units $N_h$.
Figure~\ref{f:train}(b) shows the $N_h$ dependence of $1 - K$ after 20000
updates, indicating that precision improves with an increase in $N_h$.

\section{Results}
\label{s:results}

\subsection{Interacting bosons in one dimension}

First, we apply our method to one-dimensional problems that have exact
solutions to evaluate the precision of our proposed method.
We consider the Calogero-Sutherland model~\cite{Calogero,Sutherland}, in
which bosons are confined in a harmonic potential and interact with each
other through an inverse squared potential.
The Hamiltonian for the system is given by
\begin{equation} \label{HCal}
\hat H = \sum_{i=1}^N \left( -\frac{1}{2} \frac{\partial^2}{\partial x_i^2}
+ \frac{1}{2} x_i^2 \right) + \sum_{j<k} \frac{\beta(\beta - 1)}{(x_j -
x_k)^2},
\end{equation}
where length and energy are normalized, and $\beta$ is an interaction
parameter.
The exact ground-state wave function for this Hamiltonian is available
as~\cite{Calogero}
\begin{equation} \label{exactC}
\Psi_{\rm exact} = \exp\left( -\frac{1}{2} \sum_{i=1}^N x_i^2 \right)
\prod_{j<k} |x_j - x_k|^\beta
\end{equation}
with the ground-state energy
\begin{equation}
E_{\rm exact} = \frac{N}{2} + \frac{\beta}{2} N (N - 1).
\end{equation}
The wave function in Eq.~(\ref{exactC}) satisfies the permutation symmetry
of bosons.

We prepare the initial network parameters using the method in
Sec.~\ref{s:method} with $\Psi_{\rm train}$ given in Eq.~(\ref{harmsol}),
which is the noninteracting ground state.
The interaction potential is then gradually introduced as
\begin{equation} \label{gradual}
  \hat H_{\rm int} = \min \left[ (a n)^2,
    \sum_{j<k} \frac{\beta(\beta - 1)}{(x_j - x_k)^2} \right],
\end{equation}
where $n$ is the update step and $a$ determines the ramp speed of the
potential.
Initially $n = 0$ and the interaction potential vanishes.
As the update step $n$ increases, the peak of the interaction potential
gradually rises.
This prescription can avoid the calculation breaking down due to the
sudden introduction of an infinite potential.

\begin{figure}[tb]
\includegraphics[width=8cm]{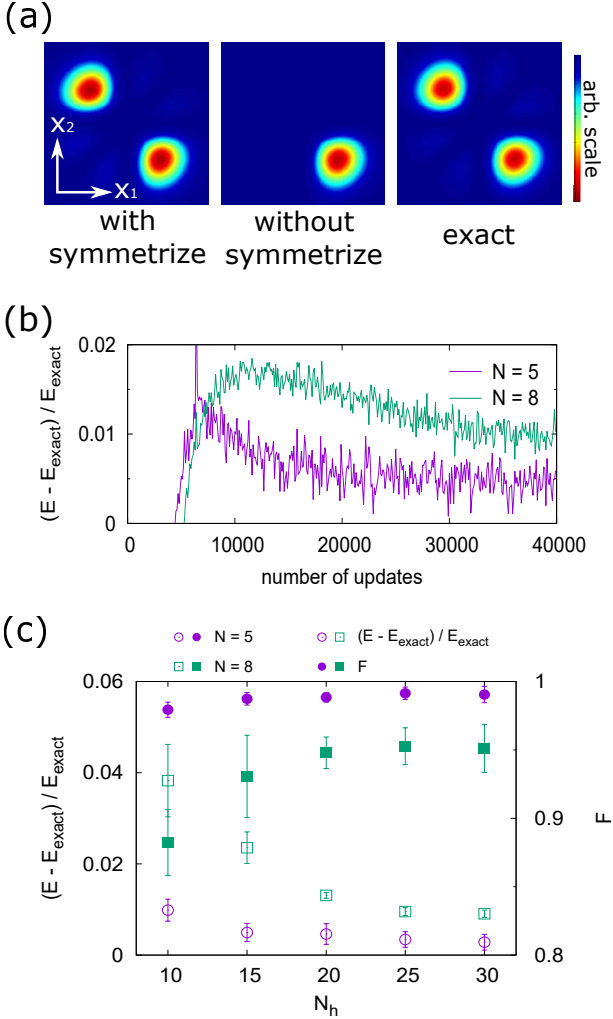}
\caption{
  (Color online) Results for the Hamiltonian in Eq.~(\ref{HCal}) with $\beta
  = 2$.
  (a) Wave function $\psi(x_1, x_2, x_3 = 0)$ for $N = 3$.
  The middle and left panels are obtained with the input functions in
  Eqs.~(\ref{simplexi}) and (\ref{permxi}), respectively, and the right
  panel shows the exact solution in Eq.~(\ref{exactC}).
  (b) Energy convergence with respect to the number of network parameter
  updates for $N = 5$ and 8.
  The number of hidden units is $N_h = 20$.
  The interaction is gradually introduced as in Eq.~(\ref{gradual}) with $a
  = 0.001$.
  The energies averaged over 100 updates are shown at intervals of 100
  updates.
  (c) Energy and fidelity after 40000 updates.
  The error bars represent standard deviations for 5 runs.
  The number of samples is $N_{\rm sample} = 10^4$ for each update in
  (a)-(c), and $N_{\rm sample} = 10^6$ to calculate the data points in (c).
}
\label{f:calogero}
\end{figure}
We first explore a simple input of particle positions as in
Eq.~(\ref{simplexi}); that is, the positions of the particles are directly
input into the neural network.
After a sufficient number of network update steps, we obtain the wave
function, as shown in the middle panel of Fig.~\ref{f:calogero}(a), which
shows $\psi(x_1, x_2, 0)$ for $N = 3$.
The permutation symmetry $x_1 \leftrightarrow x_2$ is not satisfied in this
wave function.
This is because, in the Calogero-Sutherland model, many degenerate states
exist near the ground states, such as a state satisfying
$\psi(x_1, x_2, x_3) = -\psi(x_2, x_1, x_3)$.
The linear combination of these states breaks the permutation symmetry, as
shown in the middle panel of Fig.~\ref{f:calogero}(a).
To resolve this problem, we use as an input function
\begin{equation} \label{permxi}
\bm{\xi}(\bm{X}) = (x_{i_1}, x_{i_2}, \cdots, x_{i_N}),
\end{equation}
where $x_{i_1} \leq x_{i_2} \leq \cdots \leq x_{i_N}$; that is, we input the
sorted positions into the network.
By this input method, bosonic permutation symmetry is maintained from
the definition.
The result of this input is shown in the leftmost panel of
Fig.~\ref{f:calogero}(a), which maintains the permutation symmetry, and
accords with the exact solution shown in the rightmost panel of
Fig.~\ref{f:calogero}(a).

Figure~\ref{f:calogero}(b) shows the evolution of energy with respect to
update step~\cite{Note}.
Because of the gradual increase in interaction potential as in
Eq.~(\ref{gradual}), the energy grows from $E < E_{\rm exact}$, and
stochastically converges to the final value.
Figure~\ref{f:calogero}(c) shows the energy obtained after 40000 updates.
The energy is improved as the number of hidden units $N_h$ is increased for
$N_h = 10$-20, since the representation ability of the network increases
with $N_h$.
However, the improvement saturates for $N_h \gtrsim 20$.
Such saturation is also seen in Ref.~\citen{Saito} because network
optimization becomes increasingly difficult for larger $N_h$.
Figure~\ref{f:calogero}(d) illustrates the fidelity $F$ of the wave
function, defined as
\begin{equation} \label{F}
F = \frac{\left[ \int \Psi_{\rm exact}(\bm{X}) \psi(\bm{X}) d\bm{X}
    \right]^2}{\int \Psi^2_{\rm exact}(\bm{X}) d\bm{X}
  \int \psi^2(\bm{X}) d\bm{X}}
\simeq \frac{\langle B \rangle_s^2}{\langle B^2 \rangle_s},
\end{equation}
where $B = \Psi_{\rm exact} / \psi$.
The fidelity is larger than 0.99 for $N = 5$ and larger than 0.95 for $N =
8$.
Fidelity improvement also saturates for $N_h \gtrsim 20$.

A periodic boundary condition can also be treated in our method.
Specifically, we consider the Calogero-Sutherland model in free space with a
periodic boundary condition.
The Hamiltonian for the system is given by
\begin{equation} \label{HSu}
\hat H = -\sum_{i=1}^N \frac{\partial^2}{\partial \theta_i^2}
+ \frac{\beta(\beta - 1)}{2} \sum_{j<k} \frac{1}{\sin^2\frac{\theta_j -
\theta_k}{2}},
\end{equation}
where $0 \leq \theta_i < 2\pi$.
The ground-state wave function and energy are obtained as~\cite{Sutherland}
\begin{equation}
\Psi_{\rm exact} = \prod_{j<k} \left( \sin \frac{\theta_j - \theta_k}{2}
\right)^\beta,
\end{equation}
and
\begin{equation}
E_{\rm exact} = \frac{\beta^2}{12} N(N^2 - 1).
\end{equation}

To incorporate the periodic boundary condition into the wave function
produced by the neural network, we use an input method as follows.
We first sort the positions as $\theta_{i_1} \leq \theta_{i_2} \leq \cdots
\leq \theta_{i_N}$.
If we directly input these positions into the network, a discontinuity
arises at $\theta = 0$, because the network does not know that
$\theta = 0 = 2\pi$.
To avoid this problem, we take the distance between the adjacent positions,
$\Delta_1 \equiv \theta_{i_2} - \theta_{i_1}$, $\Delta_2 \equiv \theta_{i_3}
- \theta_{i_2}$, $\cdots, \Delta_N \equiv \theta_{i_1} + 2\pi -
\theta_{i_N}$.
We also take their cyclic permutation.
The input into the network is thus
\begin{equation} \label{xis}
\bm{\xi}(\bm{X}) = (\Delta_{p}, \Delta_{p+1}, \cdots, \Delta_N, \Delta_1,
\cdots, \Delta_{p-1}),
\end{equation}
where $p$ is a random integer such that $1 \leq p \leq N$.

\begin{figure}[tb]
\includegraphics[width=8cm]{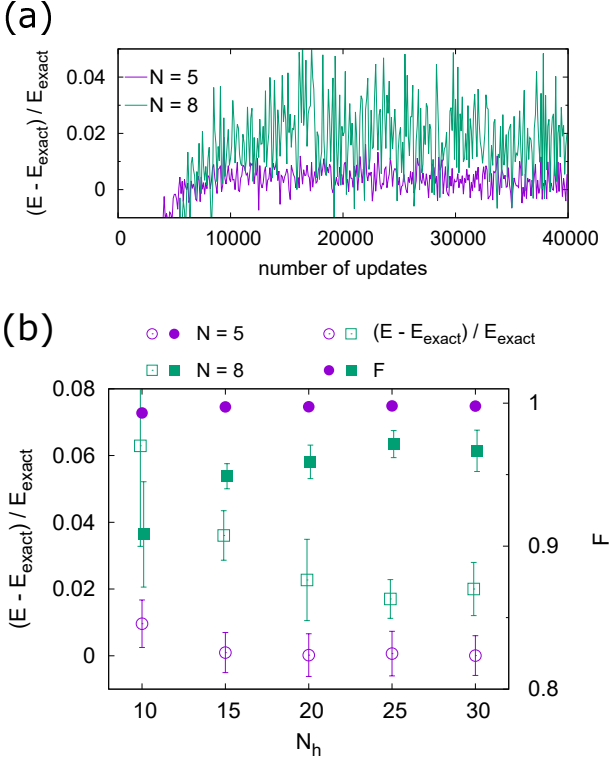}
\caption{
  (Color online) Results for the Hamiltonian in Eq.~(\ref{HSu}) with $\beta
  = 2$.
  (a) Convergence of energy with respect to the number of network parameter
  updates for $N = 5$ and 8.
  The number of hidden units is $N_h = 20$.
  The interaction is gradually introduced as in Eq.~(\ref{gradual}) with $a
  = 0.002$.
  The energies averaged over 100 updates are shown at intervals of 100
  updates.
  (b) Energy and fidelity after 40000 updates.
  The error bars represent standard deviations for 5 runs.
  The number of samples is $N_{\rm sample} = 10^4$ for each update in (a)
  and (b), and $N_{\rm sample} = 10^6$ to calculate the data points in
  (b).
}
\label{f:suther}
\end{figure}
Figure~\ref{f:suther}(a) shows the change in energy with respect to the
number of updates.
As in the previous case, we gradually increase the interaction using
Eq.~(\ref{gradual}), and the energy $E$ grows from $E < E_{\rm exact}$ and
approaches $E_{\rm exact}$.
Figure~\ref{f:suther}(b) shows the $N_h$ dependence of energy and fidelity.
Also, in this case, results improve with an increase in $N_h$ for
$N_h \lesssim 20$ and saturate for $N_h \gtrsim 20$.
The fidelity is larger than 0.99 for $N = 5$ and larger than 0.95 for $N =
8$.

\subsection{Few-body bound states in three dimensions}

We consider bosons in three-dimensional space with an attractive interaction
between particles.
As an interesting demonstration of such quantum many-body problems, we aim
to produce the Efimov bound state, which is a three-body bound state near
the resonant interaction~\cite{Efimov,Naidon}.
The Hamiltonian for the system is assumed to be
\begin{equation}
\hat H = \sum_{i=1}^N \left( -\frac{1}{2}
\frac{\partial^2}{\partial \bm{x}_i^2} + \frac{\alpha}{2} \bm{x}_i^2 \right)
- \sum_{j < k} \frac{V_0}{2 r_0^2} \exp\left[ -\frac{(\bm{x}_j -
\bm{x}_k)^2}{2 r_0^2} \right],
\end{equation}
where $\bm{x}_i$ are three-dimensional vectors of the position of the $i$th
particle, $\alpha$ is the strength of harmonic confinement, and $V_0$ and
$r_0$ are the strength and range of the Gaussian interaction, respectively.
The parameter $\alpha$ is controlled to promote bound-state formation.
We assume that interaction strength is at the first resonance, $V_0
\simeq 2.684$, at which the first two-body bound state emerges at this
interaction strength.

The Jacobi coordinate is convenient to remove the translational degree of
freedom.
We define the Jacobi coordinate $\bm{r}_i$ as
\begin{equation}
  \bm{r}_{N-j+1} = \sqrt{\frac{2j}{j+1}} \left[ \bm{x}_{j+1} - \frac{1}{j}
    \left( \bm{x}_1 + \bm{x}_2 + \cdots + \bm{x}_j \right) \right]
\end{equation}
for $j = 1, 2, \cdots, N - 1$, and 
\begin{equation}
  \bm{r}_1 = -\sqrt{\frac{2}{j(j+1)}}
  \left( \bm{x}_1 + \bm{x}_2 + \cdots + \bm{x}_N \right).
\end{equation}
Using this Jacobi coordinate, the Hamiltonian can be divided into $\hat H =
\hat H_1 + \hat H'$ with
\begin{eqnarray}
  \hat H_1 & = & -\frac{1}{N+1} \frac{\partial^2}{\partial \bm{r}_1^2}
  + \frac{N + 1}{4} \alpha \bm{r}_1^2, \\
  \hat H' & = & \sum_{i=2}^N \left( -\frac{\partial^2}{\partial \bm{r}_i^2}
  + \frac{\alpha}{4} \bm{r}_i^2 \right)
- \sum_{j < k} \frac{V_0}{2 r_0^2} \exp\left[ -\frac{(\bm{x}_j -
\bm{x}_k)^2}{2 r_0^2} \right]. \label{Hp}
\nonumber \\
\end{eqnarray}
Since the Gaussian interaction term in $\hat H'$ does not include
$\bm{r}_1$, the ground-state wave function can be written as
\begin{equation} \label{psip}
  \psi(\bm{x}_1, \bm{x}_2, \cdots, \bm{x}_N) = e^{-\frac{N+1}{4}
    \alpha^{1/2} r_1^2} \psi'(\bm{r}_2, \cdots, \bm{r}_N),
\end{equation}
where $\psi'$ is the ground state for $\hat H'$.
Thus, we seek the wave function $\psi'$ that minimizes $E' = \langle \hat H'
\rangle$.
Metropolis sampling is performed in the coordinate $\bm{X}$; i.e., we
sample $\bm{x}_1$, $\bm{x}_2$, $\cdots$, $\bm{x}_N$ (not $\bm{r}_2, \cdots,
\bm{r}_N$) in each sampling with a probability $\psi^2 = e^{-\frac{N+1}{2}
\alpha^{1/2} r_1^2} \psi^{'2}$ in Eq.~(\ref{psip}), where the network
gives $\psi' = \exp(u^{\rm out})$.

We prepare the initial state of the network parameters using
\begin{equation}
  \Psi_{\rm train} = \prod_{i=2}^N e^{-\alpha^{1/2} r_i^2/4},
\end{equation}
which is the noninteracting ground state.
The interaction is then gradually introduced as
\begin{equation}
V_0 = 2.684 \min\left( n / n_0, 1 \right),
\end{equation}
where $n$ is the number of update steps and we take $n_0 = 5000$.
Increasing the interaction gradually avoids sudden change in network
parameters which could cause the calculation to break down.

\begin{figure}[tb]
\includegraphics[width=8cm]{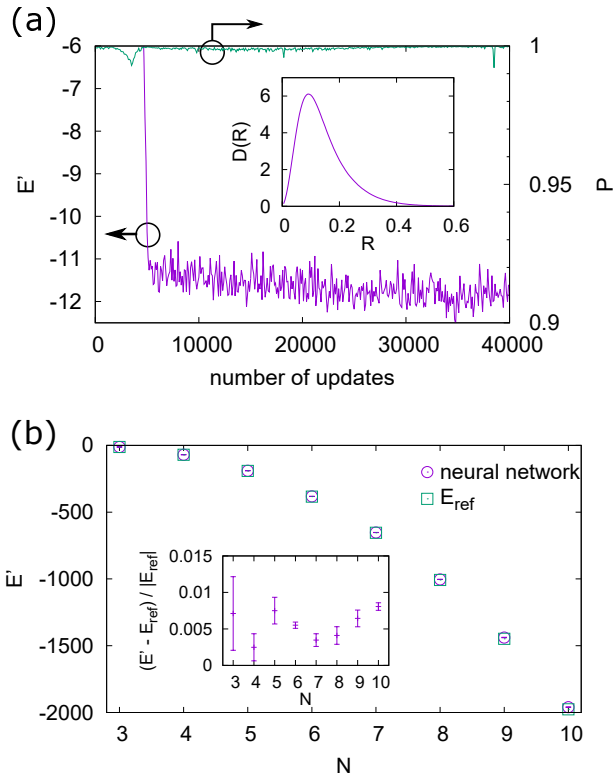}
\caption{
  (Color online) Three-dimensional bound states for the Hamiltonian in
  Eq.~(\ref{Hp}).
  (a) Changes in $E'$ and the degree of permutation symmetry $P$
  with respect to the number of network parameter updates, where $P$
  is defined in Eq.~(\ref{perm}).
  The inset shows the hyperradial distribution $D(R)$ after 40000 updates.
(b) Energies of bound states for $N = 3$-10 after 40000 updates of the
  network parameters (circles).
  The number of hidden units is $N_h = 30$ for $N \leq 7$ and $N_h = 60$ for
  $N \geq 8$.
  The squares indicate energies $E_{\rm ref}$ taken from
  Ref.~\citen{Yan}.
  The inset shows $(E' - E_{\rm ref}) / |E_{\rm ref}|$.
  The error bars represent standard deviations for 5 runs.
  The number of samples is $N_{\rm sample} = 10^4$ for each update in (a)
  and (b), and $N_{\rm sample} = 10^6$ to calculate the data points in (b).
}
\label{f:efimov}
\end{figure}
First, we consider the case of $N = 3$.
The simplest input into the network is
\begin{equation} \label{in1}
  \bm{\xi}(\bm{X}) = (r_2^{(x)}, r_2^{(y)}, r_2^{(z)},
  r_3^{(x)}, r_3^{(y)}, r_3^{(z)}),
\end{equation}
where $r_i^{(x)}$, $r_i^{(y)}$, and $r_i^{(z)}$ are $x, y$, and $z$
components of the Jacobi vector.
However, the input function in Eq.~(\ref{in1}) was found to be
inappropriate.
The calculation is unstable in many cases, and even if stability is
maintained, convergence is very slow.
This is partly because rotational degrees of freedom remain in the
coordinates in Eq.~(\ref{in1}).
We then examine the input function as
\begin{equation} \label{in2}
\bm{\xi}(\bm{X}) = (\Delta_{12}, \Delta_{13}, \Delta_{23}),
\end{equation}
where
\begin{eqnarray}
\bm{\Delta}_{12} & = & \bm{x}_1 - \bm{x}_2 = -\bm{r}_3, \nonumber \\
\bm{\Delta}_{13} & = & \bm{x}_1 - \bm{x}_3 = -\frac{1}{2} (\sqrt{3} \bm{r}_2
+ \bm{r}_3), \nonumber \\
\bm{\Delta}_{23} & = & \bm{x}_2 - \bm{x}_3 = -\frac{1}{2} (\sqrt{3} \bm{r}_2
- \bm{r}_3).
\end{eqnarray}
In this input method, the distances between particles are input into the
neural network; this takes into account not only translational symmetry
but also rotational symmetry.
Figure~\ref{f:efimov}(a) shows the change in energy with respect to
update steps.
The energy drops to $E \simeq -12$, which indicates that the bound state is
formed.
The energy is close to $-11.9$ obtained in Ref.~\citen{Yan}.
To promote formation of the bound state, we set $\alpha = 1.5$ until
10000 updates, and $\alpha = 1$ subsequently.
The inset in Fig.~\ref{f:efimov}(a) shows the hyperradial distribution
$D(R)$ of the state after convergence, where the hyperradius $R$ is
defined as $R^2= (\Delta_{12}^2 + \Delta_{13}^2 + \Delta_{23}^2) / 3$.
The particles are localized in $R \ll 1$, and hence the harmonic potential
in Eq.~(\ref{Hp}) is almost irrelevant for the binding.
In fact, a very similar result is obtained when the harmonic potential is
removed ($\alpha = 0$) after the bound state is formed.
It was also confirmed that the bound state is ``Borromean'', i.e., the bound
state is formed for $V_0 < 2.684$ for which the two-body bound state does
not exist.

In our method with the input function in Eq.~(\ref{in2}), the bosonic
permutation symmetry of the particles is not imposed explicitly.
To confirm that the obtained wave function exhibits permutation symmetry, we
introduce the overlap integral given by
\begin{eqnarray} \label{perm}
  P & = & \frac{2}{N (N + 1)} \sum_{j<k}
  \frac{\left[ \int \psi(\hat P_{jk} \bm{X}) \psi(\bm{X}) d\bm{X}
    \right]^2}{\int \psi^2(\hat P_{jk} \bm{X}) d\bm{X}
  \int \psi^2(\bm{X}) d\bm{X}} \nonumber \\
& \simeq & \frac{2}{N (N + 1)} \sum_{j<k}
\frac{\langle A_{jk} \rangle_s^2}{\langle A_{jk}^2 \rangle_s},
\end{eqnarray}
where $\hat P_{jk}$ exchanges the positions of the $j$th and $k$th
particles, and $A_{jk} = \psi(\hat P_{jk} \bm{X}) / \psi(\bm{X})$.
If the wave function exhibits permutation symmetry, $P$ becomes unity.
From Fig.~\ref{f:efimov}(a), we find that the obtained wave function
indeed acquires permutation symmetry.

Figure~\ref{f:efimov}(b) shows the energies for $N \geq 3$.
The input function is
\begin{equation}
\bm{\xi}(\bm{X}) = (\Delta_{12}, \Delta_{13}, \cdots, \Delta_{1N},
\Delta_{23}, \Delta_{24}, \cdots, \Delta_{N-1,N}),
\end{equation}
and the number of input units is $N (N - 1) / 2$.
For comparison, the energies $E_{\rm ref}$ obtained in Ref.~\citen{Yan} are
shown, in which the path-integral Monte Carlo method is used.
The energies obtained by the present method are in reasonable agreement with
those obtained by the pre-existing method.
At present, our method cannot surpass the precision of the path-integral
Monte Carlo method.
However, in our method, wave functions are obtained directly, which
facilitates the study of quantum few-body problems.

\section{Conclusions}
\label{s:conc}

We have developed a method to obtain approximate ground-state wave functions
of quantum few-body problems using artificial neural networks.
Bosons in continuous space were considered and particle positions or
their functions were input into neural networks, which were optimized to
output the desired wave functions.
The method was applied to the Calogero-Sutherland model in one-dimensional
space and the results were compared with exact solutions.
Using the appropriate input functions in Eqs.~(\ref{permxi}) and
(\ref{xis}), the bosonic permutation symmetry of the wave functions was
taken into account, and we successfully obtained the approximate
ground-state wave functions.
The method was also applied to the three-dimensional problem of bound
states with resonant interaction.
The Efimov trimer was obtained and its energy was close to that
calculated in the previous study.
Also, in this case, the appropriate input function in Eq.~(\ref{in2}) was
important to efficiently reach the ground state.
The bound states for $N > 3$ bosons were also obtained.

In the present paper, we have restricted our attention to a fully connected
neural network with a single hidden layer.
Deep neural networks with multiple hidden layers have larger representation
power, and one may posit that the use of such networks would improve
the results.
However, network optimization becomes difficult as the complexity of the
network structure increases.
In fact, we observed that the precision of results saturates as the
number of hidden units $N_h$ increases, even for the single hidden layer.
It is important to find a more efficient network structure and input method
to facilitate optimization of the network parameters.
We also need to find an appropriate way to deal with fermions.

\begin{acknowledgments}
This research was supported by JSPS KAKENHI Grant Numbers JP16K05505,
JP17K05595, JP17K05596, and JP25103007.
\end{acknowledgments}


\begin{thebibliography}{9}
\bibitem{Note}
It takes a few minutes for 20000 updates using author's workstation (Intel
Xeon Gold 6148).
\end{thebibliography}
\end{document}